\documentclass[sigconf]{acmart}

\usepackage{graphicx}
\usepackage{amsmath}
\usepackage{booktabs}
\usepackage{balance}
\usepackage{float} 
\usepackage{xcolor}
\setcitestyle{numbers,sort&compress}

\copyrightyear{2026}
\acmYear{2026}
\setcopyright{cc}
\setcctype{by}
\acmConference[KDD '26]{Proceedings of the 32nd ACM SIGKDD Conference on Knowledge Discovery and Data Mining V.2}{August 09--13, 2026}{Jeju Island, Republic of Korea}
\acmBooktitle{Proceedings of the 32nd ACM SIGKDD Conference on Knowledge Discovery and Data Mining V.2 (KDD '26), August 09--13, 2026, Jeju Island, Republic of Korea}
\acmDOI{10.1145/3770855.3818459}
\acmISBN{979-8-4007-2259-2/2026/08}



\title[PCR-CA: Parallel Codebooks with Contrastive Alignment]%
      {PCR-CA: Parallel Codebook Representations with Contrastive Alignment for Multiple-Category App Recommendation}

\author{Bin Tan}
\email{bintan@microsoft.com}
\affiliation{%
  \institution{Microsoft}
  \city{Suzhou}
  \country{China}
}

\author{Wangyao Ge}
\email{wangyaoge@microsoft.com}
\affiliation{%
  \institution{Microsoft}
  \city{Suzhou}
  \country{China}
}

\author{Yidi Wang}
\email{yidiwang@microsoft.com}
\affiliation{%
  \institution{Microsoft}
  \city{Beijing}
  \country{China}
}

\author{Xin Liu}
\email{Liu.Xin@microsoft.com}
\affiliation{%
  \institution{Microsoft}
  \city{Redmond}
  \state{WA}
  \country{USA}
}

\author{Jeff Burtoft}
\email{jeffburt@microsoft.com}
\affiliation{%
  \institution{Microsoft}
  \city{Redmond}
  \state{WA}
  \country{USA}
}

\author{Hao Fan}
\email{haofan@microsoft.com}
\affiliation{%
  \institution{Microsoft}
  \city{Suzhou}
  \country{China}
}

\author{Hui Wang}
\email{wanhui@microsoft.com}
\affiliation{%
  \institution{Microsoft}
  \city{Suzhou}
  \country{China}
}
\authornote{Corresponding author.}

\begin{document}
\begin{abstract}
Modern app store recommender systems struggle with \textit{multiple-category apps}, as traditional taxonomies fail to capture overlapping semantics, leading to suboptimal personalization. We propose \textbf{PCR-CA} (\textbf{P}arallel \textbf{C}odebook \textbf{R}epresentations with \textbf{C}ontrastive \textbf{A}lignment), an end-to-end framework for improved CTR (Click-through Rate) prediction. PCR-CA first extracts compact multimodal embeddings from app text, then introduces a \textit{Parallel Codebook VQ-AE} module that learns discrete semantic representations across multiple codebooks in parallel---unlike hierarchical residual quantization (RQ-VAE). This design enables independent encoding of diverse aspects (e.g., gameplay, art style), better modeling multiple-category semantics. To bridge semantic and collaborative signals, we employ a \textit{contrastive alignment} loss at both the user and item levels, enhancing representation learning for long-tail items. Additionally, a \textit{dual-attention fusion} mechanism combines ID-based and semantic features to capture user interests, especially for long-tail apps. Experiments on a large-scale dataset show PCR-CA achieves a +0.76\% AUC improvement over strong baselines, with +2.15\% AUC gains for long-tail apps. Online A/B testing further validates our approach, showing a +10.52\% lift in CTR and a +16.30\% improvement in CVR (Conversion Rate), demonstrating PCR-CA's effectiveness in real-world deployment. The new framework has now been fully deployed on the Microsoft Store.
\end{abstract}

\maketitle

 \section{Introduction}
 \begin{figure}[t]
\centering
\includegraphics[width=0.5\textwidth]{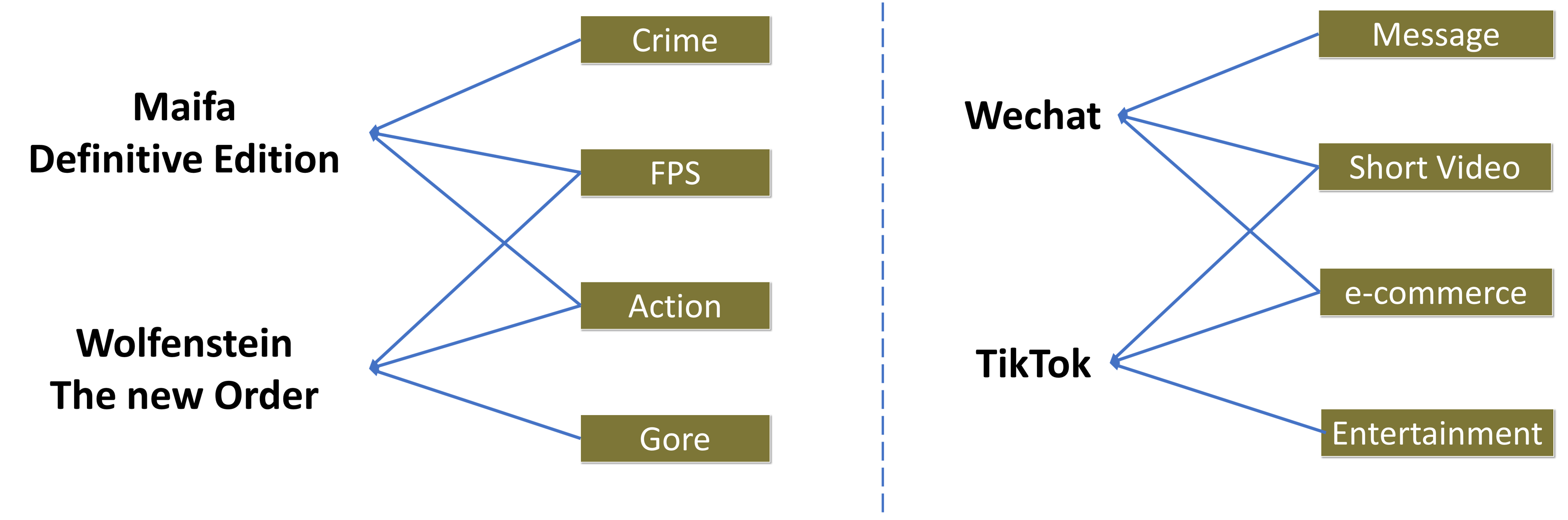}
\vspace{-2mm}
\caption{Example of multiple-category.}
\label{fig:category}
\vspace{-4mm}
\end{figure}
Recommender systems in app stores aim to match users with useful apps, but a key challenge lies in their \textbf{multiple-category} nature. Modern apps often span several categories, defying traditional tree-structured taxonomies. As illustrated in Figure~\ref{fig:category}, we compare two games (games are subcategory of apps)---\textit{Mafia} and \textit{Wolfenstein}, two famous no-game apps---\textit{WeChat} and \textit{TikTok}. Through gameplay, style and features analysis, we identify four overlapping categories. \textit{Mafia} belongs to Crime, FPS (First-Person Shooter), and Action, while Wolfenstein fits FPS, Action, and Gore. Similarly, \textit{WeChat} is primarily a messaging app and \textit{TikTok} an Entertainment app, yet both include Short Video and e-commerce features. These examples highlight the complex semantic relationships between apps and the inadequacy of single-label assignments. Traditional CTR models often treat category labels as fixed categorical features---e.g., labeling \textit{Wolfenstein} simply as ``FPS''---which oversimplifies app semantics and leads to inaccurate feature representation and suboptimal recommendations. Capturing such nuanced, multi-dimensional relationships requires more expressive modeling beyond conventional approaches.

\begin{figure}[h]
\centering
\includegraphics[width=0.5\textwidth]{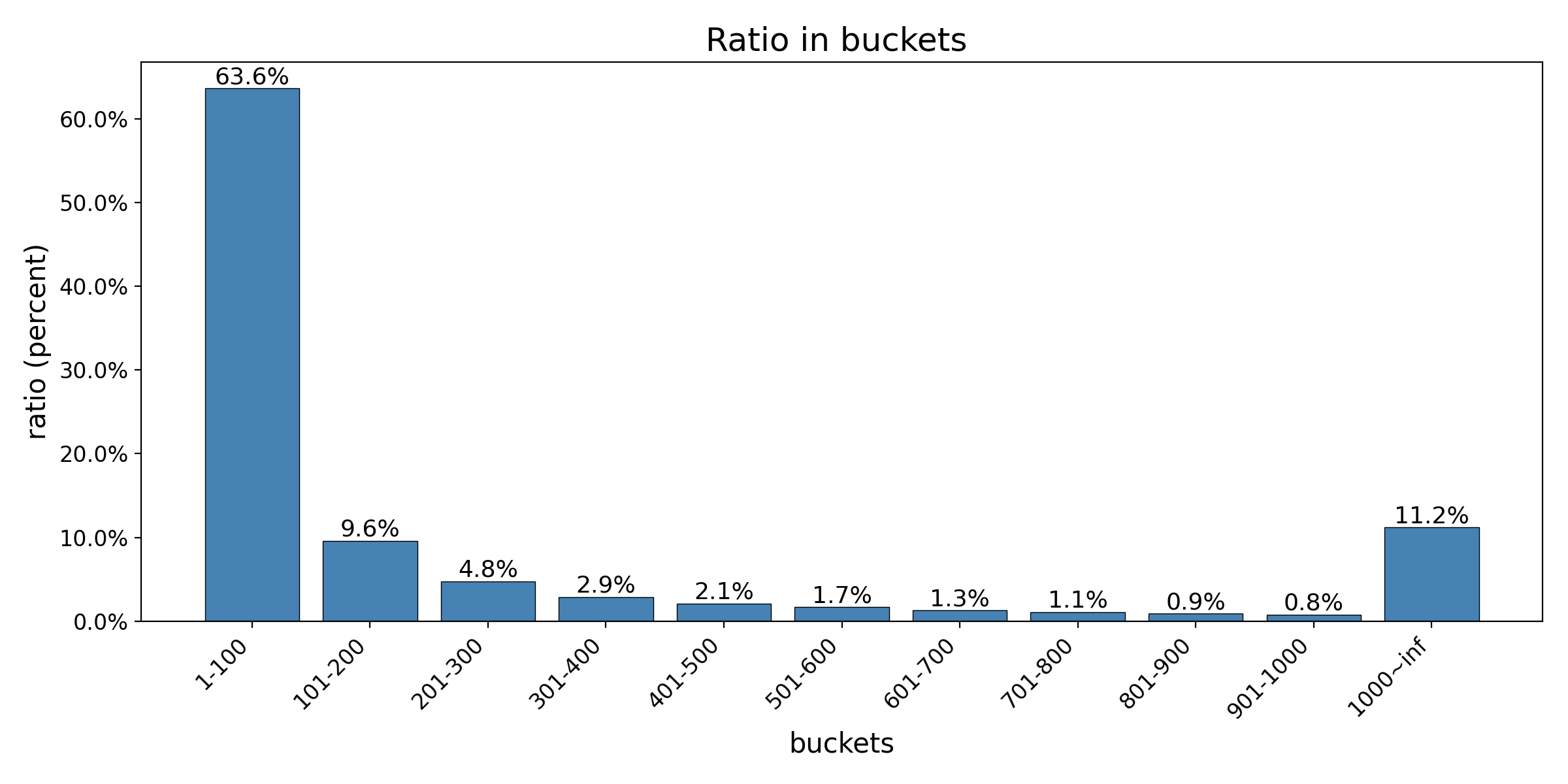}
\vspace{-2mm}
\caption{Click distribution by popularity buckets (100 apps per bucket).}
\label{fig:clickdist}
\vspace{-4mm}
\end{figure}

Another critical challenge is the \textbf{head--tail imbalance}: a small set of highly popular apps dominates user interactions, while the vast majority of niche apps receives minimal feedback, reinforcing popularity bias and reducing recommendation diversity. This phenomenon is largely driven by curated ``Best-Selling'' and ``Top-Free'' collections, combined with users' strong tendency to actively search for headline applications. Figure~\ref{fig:clickdist} illustrates the skewed click distribution across popularity buckets. In our dataset of over 80K apps, the top 100 apps alone account for 63.6\% of all clicks, reflecting an extreme head concentration effect. Consequently, the remaining 99\% of apps suffer from severe data sparsity, leaving CTR models with insufficient positive samples to learn reliable patterns for the long tail, which significantly degrades prediction accuracy for less popular items.

To address these, many models incorporate content (text, images, categories) alongside IDs. Early works like Factorization Machines and deep CTR models (Wide\&Deep, DeepFM)~\cite{cheng2016wide,guo2017deepfm} fuse metadata with embeddings. Later methods use attention or multi-layer fusion; large-scale systems (e.g., YouTube) combine diverse features. In app recommendation, multi-view and collaborative deep learning~\cite{elkahky2015mv,wang2015cdl} leverage text and tags; VBPR~\cite{he2016vbpr} adds images. However, raw high-dimensional embeddings can introduce noise and latency.
Recent studies compress content into discrete latent factors for better integration with collaborative signals, inspired by \textbf{Vector Quantization (VQ)}~\cite{oord2017vqvae}. DST~\cite{liu2024tokenization}, CoST~\cite{zhu2024cost}, and SaviorRec~\cite{yao2025saviorrec} show discrete codes improve generalization and long tail. Yet residual VQ imposes a hierarchy (broad class first, details later)~\cite{zeghidour2021soundstream}, unsuitable for multiple-category apps.

On the user side, attention mechanisms~\cite{bahdanau2015neural} effectively weight historical items for CTR. While transformer-based variants exist, they add complexity without clear gains for short sequences. A simpler two-stream attention---separating ID and content signals---can be more robust.

\textbf{Our Approach:} We propose \textbf{PCR-CA}, which (1) learns \textit{parallel codebooks} for content, avoiding hierarchical constraints; (2) applies \textit{contrastive alignment} to bridge semantic and ID spaces; and (3) uses a \textit{dual-attention fusion} to combine ID- and content-based relevance. This design improves multiple-category understanding and long-tail performance with minimal overhead.

\section{Related Work}

\subsection{Multimodal and Content-Aware Recommendation}
Incorporating rich content features (text, images, etc.) into recommenders is a long-standing strategy to address data sparsity and cold-start issues~\cite{zhang2019survey,deldjoo2020multimedia}. Early works combined content with collaborative filtering by feature augmentation. For example, Factorization Machines~\cite{rendle2010fm} integrated item metadata (category, tags) into matrix factorization by learning feature interaction weights. Google's \textit{Wide\&Deep} framework~\cite{cheng2016wide} blended a linear model (wide part, using hand-crafted features) with a deep neural network (embedding IDs), improving app recommendations by capturing both memorization and generalization. \citet{guo2017deepfm} proposed \textit{DeepFM}, which replaces the wide part with an FM component and shares embedding parameters with a deep component---thus automatically learning high-order feature interactions. Deep \& Cross Network (DCN)~\cite{wang2017deep} explicitly models cross feature interactions via cross layers, and AutoInt~\cite{song2019autoint} uses multi-head self-attention to capture high-order feature interactions.

Beyond manually engineered features, deep learning enabled end-to-end learning of content representations. \citet{elkahky2015mv} used a multi-view deep neural network for cross-domain recommendations, where item content from different domains was used to learn a shared user embedding. \citet{wang2015cdl} proposed Collaborative Deep Learning (CDL), an autoencoder that processes item text (e.g., app descriptions) in tandem with probabilistic matrix factorization---using content to regularize latent factors. These models showed that item text or tags can significantly boost recommendations when interaction data are sparse.

Recent research has looked at more complex fusion and pre-training techniques. MMDIN~\cite{yang2021mmdin} extended targeted attention by using multiple heads that focus on different modalities---e.g., one head attending to image features in the user's history, another to text features. Graph neural networks have also been applied: \citet{he2021hypergraph} constructed a hypergraph connecting users with the multiple content modalities of items they interacted with, and performed hypergraph convolution to propagate information. \textit{SaviorRec}~\cite{yao2025saviorrec} fine-tuned a CLIP-like multimodal embedding model on user behavior data, then used it to generate item embeddings.

While these methods advanced the state of the art, using raw high-dimensional content features can still be problematic. As noted by \citet{liu2024tokenization}, directly concatenating a 768-d BERT embedding of an item may introduce many parameters without clear gains. This motivates our approach of learning a compressed, discrete representation of content that the model can interpret more easily.

\subsection{Discrete Representation Learning in Recommenders}
Representing entities with discrete codes has gained traction due to efficiency and potential regularization benefits. Storing a few code indices per item is far more compact than storing a dense vector, and discrete codes can capture combinatorial structure.

One line of work is semantic hashing, mapping items or documents to binary or discrete codes such that similar items have similar codes. Early attempts in recommendation include \citet{salakhutdinov2009semantic}, who used RBMs to learn binary codes for items. Modern deep versions include DST~\cite{liu2024tokenization}, which learned binary codes for user and item ID embeddings. However, binary codes may be too restrictive in capacity.

Vector quantization allows more than two values per code. \citet{oord2017vqvae} introduced a method to embed inputs in a continuous space and then quantize to the nearest code in a learned codebook, training with a combination of reconstruction and codebook update losses. This has been highly successful in compressing images, audio, etc. \citet{zhu2024cost} and \citet{yao2025saviorrec} both use a variant called Residual VQ (RQ-VAE) which stacks multiple quantization layers. RQ-VAE provides multiple codebooks: the first quantizes the raw input, then the residual error is quantized by a second codebook, and so on. In recommenders, RQ-VAE has been used to encode an item's multimodal feature vector into a sequence of codes. \citet{zhu2024cost} found that using two residual codes per item maintained good representation quality while reducing item embedding storage by orders of magnitude. Similarly, \citet{yao2025saviorrec} reported that replacing a 64-d continuous embedding with two 8-d code indices produced a meaningful semantic representation that improved cold-start performance.

\textbf{Relation to Product Quantization and RQ-VAE.} Product Quantization (PQ)~\cite{jegou2011product} decomposes a high-dimensional vector space into a Cartesian product of subspaces, each quantized independently. While PQ aims primarily at compression efficiency through \textit{space decomposition}, our parallel codebooks extend this idea to \textit{semantic decomposition}: each codebook is trained to capture a distinct semantic aspect (e.g., gameplay, art style) rather than an arbitrary subspace. Residual VQ (RQ-VAE)~\cite{zeghidour2021soundstream} stacks codebooks hierarchically, enforcing a coarse-to-fine inductive bias. This hierarchical structure assumes that semantics can be neatly layered from broad to detailed, which is natural for audio or images but less suitable for apps whose categories overlap without clear hierarchy. SaviorRec~\cite{yao2025saviorrec} also employs RQ-VAE and contrastive alignment; however, its residual design remains hierarchical. PCR-CA removes this ordering constraint by using parallel, independent codebooks, which we empirically show better captures multi-faceted app semantics (see Section~\ref{sec:experiments}).

Another important aspect is aligning these content-derived codes with collaborative information. Wei et al.~\cite{wei2021contrastive} and Pan et al.~\cite{pan2021click} also propose contrastive losses to align content-based and ID-based item embeddings in CTR prediction. \citet{yao2025saviorrec} addressed this by merging the semantic code into the item embedding and adding a contrastive loss between the content-based and ID-based embeddings. We extend this idea significantly: our contrastive alignment aligns representations at both the item and user levels and uses careful sampling to avoid misleading negatives. Other works have also explored aligning content and ID spaces: \citet{lin2025unified} unified two embedding spaces via a shared decoder, and \citet{zheng2025semanticID} encouraged consistency between old and new item IDs via distillation, finding that semantic IDs improved temporal stability. Our method directly optimizes alignment during training using an InfoNCE-style loss~\cite{oord2018infonce}, which effectively enforces consistency between the content and collaborative spaces.

\subsection{Attention Mechanisms for User Behavior}
Neural attention originated in sequence-to-sequence models for machine translation~\cite{bahdanau2015neural}, and it was first applied to recommendation by using attention to pinpoint relevant items in a user's history for a target item. In personalized recommendation, such \textit{attention networks} assign a weight to each past item, reflecting its importance in predicting the user's interest in the current item.
Attention-based models further inspire our design. DIN~\cite{zhou2018din} applies target-aware attention to weight user behaviors by relevance. MMDIN~\cite{yang2021mmdin} extends this with multi-head attention to fuse ID and multi-modal features. MIM~\cite{yan2025mim} introduces multi-perspective attention to capture dynamic preferences and bias factors.

In our model, we incorporate a similar attention mechanism, but with a twist to handle two feature modalities (ID and content). One could consider more complex schemes like multi-head attention mixing modalities or even a full transformer encoder over item sequences~\cite{kang2018sasrec}. In multi-task learning, \citet{ma2018mmoe} used separate expert subnetworks and a gating mechanism to combine them, ensuring each expert specializes. Analogously, we want one ``expert'' to focus on ID-based patterns and another to focus on semantic patterns. We considered gating mechanisms but found that a simple average was effective and more stable. Indeed, bagging ensembles~\cite{breiman1996bagging} suggest that averaging predictors can reduce variance. Similarly, Zhao et al.~\cite{zhao2023bootstrapping} apply a bootstrapping contrastive learning scheme to mitigate long-tail representation bias in cold-start music recommendation.

\section{Methodology}
\begin{figure*}[t]
\centering
\includegraphics[width=\textwidth]{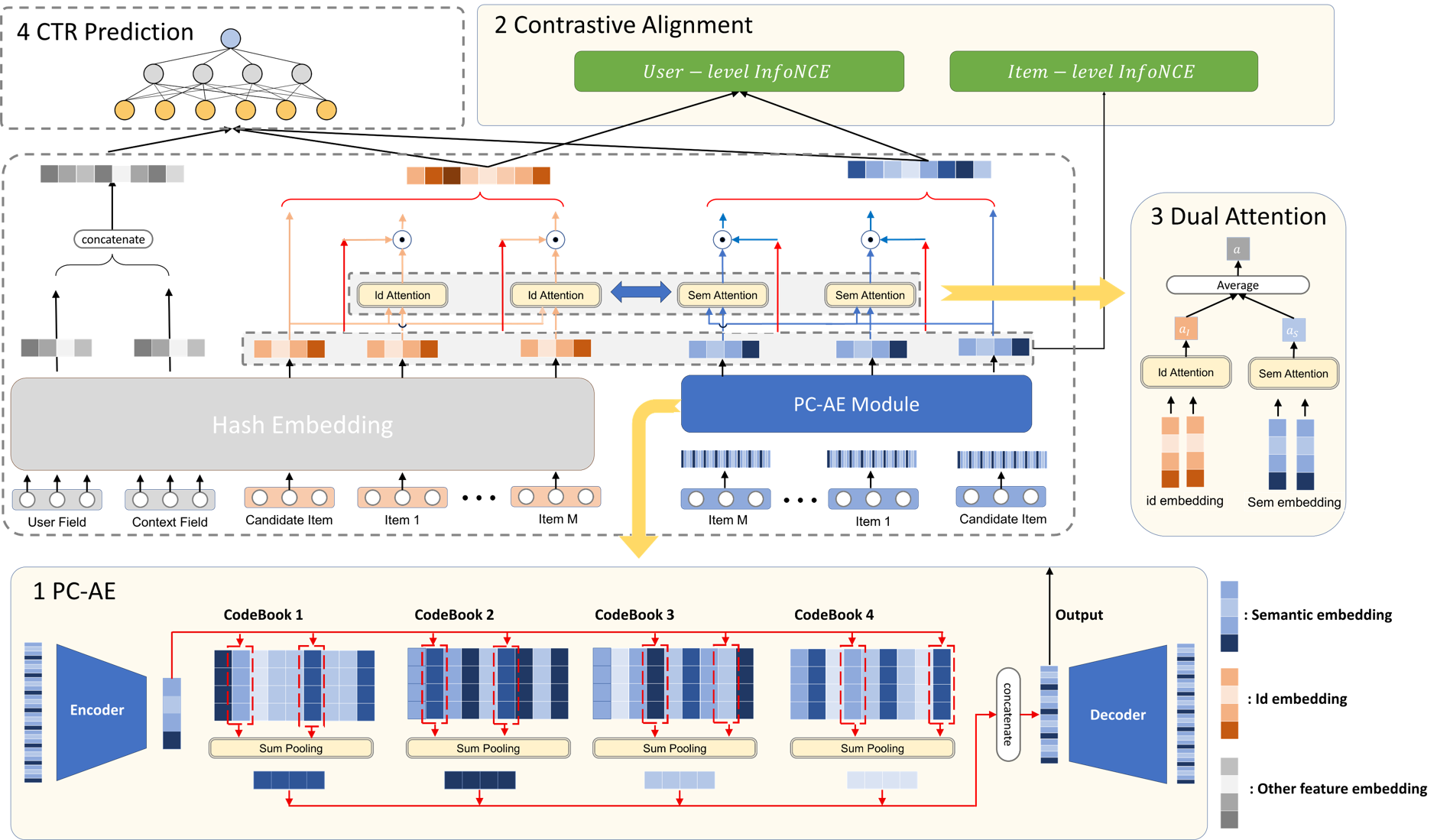}
\vspace{-2mm}
\caption{Overview of the PCR-CA framework which consists of four components: (1) PC-AE, ; (2) contrastive alignment module,; (3) dual attention; and (4) the CTR prediction network.}
\label{fig:arch}
\vspace{-3mm}
\end{figure*}
\label{sec:method}
Figure~\ref{fig:arch} illustrates the overall architecture of \textbf{PCR-CA}, which consists of four components:
(1) \textit{Parallel Codebook Auto-Encoder (PC-AE)}, which compresses high-dimensional multi-modal embeddings into discrete code vectors and reconstructs them, jointly trained with the main CTR model using reconstruction, alignment, and CTR losses;
(2) \textit{Contrastive Alignment}, which computes alignment losses on both user and item levels, where user representations are aggregated from behavior sequences and item representations include both ID and semantic embeddings;
(3) \textit{Dual-Attention Fusion}, which trains two attention networks for ID and semantic embeddings and integrates them by averaging their outputs over user history;
and (4) the \textit{CTR Prediction Network}, a standard embedding-plus-MLP architecture with attention, optimized using cross-entropy loss.

\subsection{Multimodal Embedding Extraction}
Each app may have multiple content modalities. We use textual features (title and description) as they are universally available and rich in semantics. We concatenate an app's title and description into a single text string and feed it through a pre-trained large language text embedding model (here we used Microsoft text-embedding model). We take the output as the initial item vector (256 dimensions), denoted $\mathbf{e}_i$ for item $i$, to serve as input to the codebook module.

We keep the text encoder fixed to leverage large-scale language training and avoid expensive fine-tuning on our corpus. In preliminary experiments, we found that even without fine-tuning the text encoder, the subsequent learning can adapt the representations well. One could fine-tune on domain-specific data for potentially better alignment~\cite{yao2025saviorrec}, but we did not find it necessary.

\textbf{We deliberately exclude images} because, in app recommendation, images (especially game covers) are often designed to be visually provocative and may encourage click-bait behavior that hurts long-term user satisfaction and store atmosphere. Moreover, our preliminary analysis showed that text embeddings already encode sufficient world knowledge via the LLM (e.g., recognizing \textit{DeepSeek} as an AI app rather than a generic search app), making images redundant for semantic understanding while introducing unnecessary serving cost.

\subsection{Parallel Codebook Auto-Encoder (PC-AE)}
The PC-AE compresses the continuous item embedding $\mathbf{e}_i$ into a set of discrete code vectors. It consists of an encoder network, multiple codebooks, and a decoder network. This extends the classic VQ-VAE into a multi-codebook setting.

\textbf{Encoder:} We apply a two-layer MLP to the 256-dimensional input $\mathbf{e}_i$ to obtain a 16-dimensional latent vector $\mathbf{z}_{enc,i}$. A LayerNorm layer is added after the MLP to stabilize training. This encoder learns to emphasize aspects of $\mathbf{e}_i$ that are important for reconstruction and CTR prediction.

\textbf{Quantization:} The latent vector $\mathbf{z}_{enc,i}$ is mapped to discrete codes by selecting the top-$K$ nearest code vectors from each of the $N$ codebooks in Euclidean space:
\[
\mathbf{q}^{(j)}_i = \sum_{k \in \mathrm{TopK}(\mathbf{z}_{enc,i}, C^{(j)})} \mathbf{c}^{(j)}_k,
\]
with $K=2$ in our implementation. Allowing multiple codes per book enables each codebook to capture multiple sub-aspects. For example, one codebook might encode ``gameplay,'' and an app could activate both \texttt{FPS} and \texttt{Action} codes if it mixes those gameplays.

After quantization, we concatenate the $N$ code vectors to form $\mathbf{s}_i = [\mathbf{q}^{(1)}_i \parallel \cdots \parallel \mathbf{q}^{(N)}_i]$, which serves as the learned \textbf{semantic embedding} for item $i$. The embedding is determined entirely by the selected code indices, which can be stored compactly as $(c^{(1)}, \dots, c^{(N)})$.

\textbf{Decoder:} We feed $\mathbf{s}_i$ into a two-layer MLP to reconstruct the original input embedding $\mathbf{e}_i$, producing $\hat{\mathbf{e}}_i$. The reconstruction loss is:
\[
L_{\text{recon}} = \|\hat{\mathbf{e}}_i - \mathbf{e}_i\|_2^2.
\]

Additionally, we apply VQ commitment losses~\cite{oord2017vqvae} for each codebook $j$ to optimize training stability and improve codebook utilization:
\[
L_{\text{vq}}^{(j)} = \|\mathbf{z}_{enc,i} - \text{sg}(\mathbf{q}^{(j)}_i)\|_2^2 + \beta \|\mathbf{q}^{(j)}_i - \text{sg}(\mathbf{z}_{enc,i})\|_2^2,
\]
where $\text{sg}$ is the stop-gradient operator and $\beta=0.25$. These terms encourage encoder outputs to stay near selected codes and update code vectors toward encoder outputs, preventing codebook collapse.

\textbf{Training Strategy.} The PC-AE module is trained in three stages to ensure stable codebook learning, especially for long-tail items. \textbf{Stage 1:} We train only the encoder--decoder (autoencoder) on all 80K apps for 100 epochs, learning a compact latent space without quantization. \textbf{Stage 2:} We add the $N$ parallel codebooks and continue training for another 100 epochs to establish reliable semantic codes across the entire catalog. \textbf{Stage 3:} We perform end-to-end joint training with the CTR task on exposure/click samples for 1 epoch. The final stage is limited to one epoch because multi-epoch training on large-scale click logs commonly leads to overfitting in industrial recommender systems. This staged procedure prevents head-item dominance from biasing codebook initialization and guarantees that long-tail apps receive adequate semantic learning before CTR optimization begins.

\subsection{Contrastive Alignment Module}
Even with joint training, there may remain a gap between the space of $\mathbf{s}_i$ (the code-based embedding of item $i$) and $\mathbf{v}_i$ (the ID embedding of item $i$ from collaborative data). If not aligned, the model might treat these as distinct features. We introduce contrastive learning objectives to explicitly align them. We design two contrastive tasks:

\textbf{Item-level alignment:} We treat an item's semantic embedding and its ID embedding as a positive pair. Additionally, if two items $p$ and $q$ were co-installed by the same user within a \textbf{7-day window}, or co-clicked within a \textbf{24-hour window}, we treat $\mathbf{s}_p$ with $\mathbf{v}_q$ (and vice versa) as positive as well. We selected the 7-day install window by testing $\{7, 14, 30\}$ days: longer windows generate too many positives and dilute specificity, while 7 days achieved the best offline AUC. The 24-hour click window is used because click behavior is much denser than install behavior; restricting installs to 24 hours would yield too few positive pairs. Intuitively, this shares semantic information among similar items, not just identical items, making the alignment more robust. All other unrelated item pairs in a batch are considered negatives.

We use an InfoNCE loss~\cite{oord2018infonce}. For a given item $i$, the item-level contrastive loss is:
\[
L_{\text{item}}^{(i)} = -\log \frac{\sum_{j \in \mathcal{P}(i)} \exp(\text{sim}(\mathbf{s}_i, \mathbf{v}_j) / \tau)}{\sum_{j \in \mathcal{P}(i) \cup \mathcal{N}(i)} \exp(\text{sim}(\mathbf{s}_i, \mathbf{v}_j)/\tau)},
\]
where $\mathcal{P}(i)$ is the set of positive matches for $i$ (including $i$ itself, and possibly a few related items), $\mathcal{N}(i)$ is the set of negatives (other items in the batch not in $\mathcal{P}(i)$), and $\text{sim}(x,y)$ denotes cosine similarity. We similarly define a counterpart loss treating $\mathbf{v}_i$ as the anchor and $\mathbf{s}_j$ as targets (symmetric). In practice we compute a combined loss over both anchor types.

\textbf{User-level alignment:} We also align user representations. We define a user's ID-based representation $\mathbf{u}^{ID}_u$ as the mean of ID embeddings of their history and a user's content-based representation $\mathbf{u}^{SEM}_u$ as the mean of semantic embeddings of their history. These two should be close for the same user. We treat $(\mathbf{u}^{ID}_u, \mathbf{u}^{SEM}_u)$ as a positive pair for each user $u$, and $(\mathbf{u}^{ID}_u, \mathbf{u}^{SEM}_v)$ for $u \neq v$ as negatives. A similar InfoNCE loss is:
\[
L_{\text{user}}^{(u)} = -\log \frac{\exp(\text{sim}(\mathbf{u}^{ID}_u, \mathbf{u}^{SEM}_u)/\tau)}{\sum_{v} \exp(\text{sim}(\mathbf{u}^{ID}_u, \mathbf{u}^{SEM}_v)/\tau)}.
\]

The total contrastive alignment loss is $L_{\text{align}} = \lambda_{\text{i}} \sum_i L_{\text{item}}^{(i)} + \lambda_{\text{u}} \sum_u L_{\text{user}}^{(u)}$, where we set $\lambda_{\text{i}}=1.0$ and $\lambda_{\text{u}}=1.0$ to balance the two terms. The temperature is $\tau=0.07$. The total alignment weight is $\lambda_{\text{align}}=0.01$, chosen empirically because the raw alignment loss is roughly two orders of magnitude larger than the CTR loss, and a smaller weight prevents it from dominating optimization.

This alignment yields an embedding space where, for example, a new app's semantic code embedding lies near the ID embeddings of similar existing apps. Thus, when the new app appears, the model can relate it to users and items as if it had collaborative data, even though it doesn't. The contrastive learning effectively transfers collaborative knowledge into the content-based embedding space.

\subsection{Dual-Attention Fusion for User History}
Users often have a sequence of past interactions. We adopt an attention mechanism to select relevant parts of a user's history when scoring a target item.

Formally, let $U$ be a user with history $H_U = [i_1, i_2, \dots, i_T]$, which is a list of item IDs the user interacted with in chronological order or an unordered set of past items. We evaluate a candidate item $c$ for user $U$. We compute:

 \textbf{ID-based attention:} For each history item $i_j \in H_U$, we compute a relevance score to candidate $c$ using only ID embeddings. Let $\mathbf{v}_{i_j}$ be the ID embedding of $i_j$ and $\mathbf{v}_c$ that of $c$. We feed $(\mathbf{v}_{i_j}, \mathbf{v}_c, \mathbf{v}_{i_j} \circ \mathbf{v}_c)$ into a two layer MLP that outputs a score $e_{j}^{ID}$. (The $\circ$ denotes elementwise multiplication, capturing interaction features as in \cite{zhou2018din}.) We then compute an attention weight $\alpha_j^{ID} = \frac{\exp(e_j^{ID})}{\sum_{k=1}^T \exp(e_k^{ID})}$.

 \textbf{Semantic-based attention:} Similarly, let $\mathbf{s}_{i_j}$ be the semantic embedding (code-based) of $i_j$, and $\mathbf{s}_c$ that of $c$. We compute $e_j^{SEM}$ in the same way but using $(\mathbf{s}_{i_j}, \mathbf{s}_c, \mathbf{s}_{i_j} \circ \mathbf{s}_c)$ as input to another attention network. This yields $\alpha_j^{SEM}$.

We then combine the two attention distributions using an aggregation function $g(\alpha_j^{ID}, \alpha_j^{SEM})$. In our implementation, we adopt a simple average, i.e., $g(a,b) = (a+b)/2$, because this bagging-style approach has been widely shown in machine learning competitions and industrial systems to be one of the simplest yet most effective strategies for improving robustness and reducing variance.

Using these final weights, we create aggregated user history embeddings in each space:
\[
\mathbf{h}_U^{ID} = \sum_{j=1}^T \alpha_j \, \mathbf{v}_{i_j}, \qquad
\mathbf{h}_U^{SEM} = \sum_{j=1}^T \alpha_j \, \mathbf{s}_{i_j}.
\]
Intuitively, $\mathbf{h}_U^{ID}$ summarizes the history items most relevant to $c$ in collaborative terms, while $\mathbf{h}_U^{SEM}$ summarizes those most relevant in content.

Finally, we feed the following into the prediction MLP: the candidate's ID embedding $\mathbf{v}_c$, the candidate's semantic embedding $\mathbf{s}_c$, the aggregated history $\mathbf{h}_U^{ID}$, $\mathbf{h}_U^{SEM}$, and optionally the user's own ID embedding (if used). The output is $\hat{y}_{Uc}$, the predicted probability that user $U$ will click/install item $c$.

We train the model to minimize the standard binary cross-entropy loss $L_{\text{CTR}} = -[y_{Uc}\log \hat{y}_{Uc} + (1-y_{Uc})\log(1-\hat{y}_{Uc})]$ for each training instance $(U,c)$. The total loss is:
\[
L = L_{\text{CTR}} + \lambda_{\text{rec}} L_{\text{recon}} + \lambda_{\text{vq}} \sum_{j=1}^N L_{\text{vq}}^{(j)} + \lambda_{\text{align}} L_{\text{align}},
\]
with hyperparameters to balance components. We typically set $\lambda_{\text{rec}} \approx 1$ and choose $\lambda_{\text{align}}=0.01$, since the absolute value of the alignment loss is two orders of magnitude larger than the CTR loss, and a smaller weight prevents it from dominating optimization. The commitment loss weight $\lambda_{\text{vq}}$ is 0.25.

 Inspired by bagging in ensemble learning, our dual-attention design fuses two complementary information sources---ID-based and content-based attention---by averaging their weights. Although the offline AUC gain over a single mixed-attention is modest (see Table~\ref{tab:ablation-attn}), this design consistently improves training stability and generalization. App-store user behavior is sparser than e-commerce or short-video domains; simply increasing fitting capacity does not necessarily improve online performance. Instead, the bagging-style dual-stream design acts as a simple but effective mechanism to transfer interest patterns learned from one user to another with similar sparse behavior sequences, thereby improving robustness.

\section{Experiments}
\label{sec:experiments}
We evaluate \textbf{PCR-CA} on a real-world app recommendation scenario and compare it with state-of-the-art baselines. We focus on: (1) Does PCR-CA improve overall CTR prediction and ranking quality? (2) How does it perform across different popularity segments, especially for long-tail apps? (3) How do individual components (parallel codebooks, contrastive loss, dual attention) contribute?

\subsection{Dataset and Experimental Setup}
\begin{table}
\centering
\caption{The statistic of Experimental dataset.}
\label{tab:dataset}
\begin{tabular}{l r}
\toprule
\textbf{Statistic} & \textbf{Value} \\
\midrule
Training interactions & 340M \\
Validation interactions & 10M \\
Positive rate (Click) & 6.148\% \\
Distinct apps & 80{,}686 \\
\bottomrule
\end{tabular}
\vspace{-4mm}
\end{table}

\textbf{Dataset:} We collected interaction logs from the Microsoft Store, consisting of app impression records and whether the user installed or clicked the app over several months. The training set contains 340 million interactions, and the validation set includes the last 10 million samples, split chronologically to simulate a realistic temporal scenario, naturally introducing new apps and users unseen during training. The positive (click) rate is 6.148\%, and the dataset covers approximately 80K distinct apps. For each user in training, we include up to their past 50 interactions as the user history for the attention models.

\textbf{Baselines:} We adopt the model currently deployed in the Microsoft Store online service and several classic multimodal-feature-based recommender systems as baselines; for fairness, all multimodal features used are generated by the same LLM text-embedding model.
\begin{itemize}
\item \textbf{Base:} A strong production-style CTR model currently deployed in the Microsoft Store, using only user feature and item ID embeddings, with an attention mechanism over the user's item history .
\item \textbf{MV-DNN}~\cite{elkahky2015mv} A multimodal baseline that directly concatenates raw textual embeddings (256-d) of items alongside ID embeddings. The content vector is projected to 16-d and concatenated with ID features. This lacks quantization or alignment, serving as a variant of PCR-CA without codebooks.
\item \textbf{MMDIN}~\cite{yang2021mmdin} A multimodal deep interest network that applies attention over both ID and content embeddings. It uses multi-head ResNet to predict CTR. Unlike PCR-CA, it does not use quantization and alignment.
\item \textbf{MIM}~\cite{yan2025mim} This model introduces multimodal features and employs cross-modal attention across modalities to improve performance; however, it does not use quantization.
\item \textbf{SaviorRec}~\cite{yao2025saviorrec} A recent method that uses an RQ-VAE and contrastive alignment, with a Bi-Directional target attention fusion. We reimplemented it as described for comparison.
\item \textbf{PCR-CA (ours):} Our model with 4 parallel codebooks ($M=32$, $d=16$, $K=2$), user-item level contrastive alignment, and the bagging dual attention.
\end{itemize}

For fair capacity comparison, all models use 16-dimensional ID embeddings. For content features, we use a 256-d pre-trained text embedding. The prediction network for all models is a 4-layer MLP with ReLU activations on the concatenated features (user representation, item representations, etc.).

We set the codebook embedding size to 16 in order to align with the ID embeddings, and employed four codebooks, each with a dimension of 32. This choice is motivated by the nature of app ecosystems in the store: unlike short-video or e-commerce domains where items are highly diverse, apps exhibit relatively lower heterogeneity. Using excessively large embedding sizes would introduce unnecessary noise rather than improve representation capacity. We systematically explored $N \in \{2,3,4,5,6\}$ and $K \in \{2,3,4,5\}$, and found that $N=4, K=2$ achieved the best trade-off between expressiveness and training stability. Larger $K$ (e.g., 4 or 5) often caused convergence difficulties, while $N>4$ did not yield further gains.

We optimize the models with Adam at a learning-rate of 1e-3 on a parameter-server cluster of 12 machines (8 CPU cores and 56 GB RAM each). Training is stopped after a single epoch of CTR joint training, which suffices for convergence and prevents overfitting. As detailed in Section~\ref{sec:method}, the PC-AE undergoes a three-stage warm-up (100 epochs autoencoder + 100 epochs codebook + 1 epoch joint CTR) to ensure stable semantic learning for long-tail items.

\textbf{Metrics:} We report AUC (area under ROC) as the main metric. We also break down AUC by item popularity bucket (based on app popularity: [0,50), [50,100), [100,200),[200,300),[300,400),[400,500), [500,$+\infty$) to assess performance on different segments. Additionally, we note relative improvements versus the base model.

\subsection{Overall Performance}

\begin{table*}[t]
\centering
\begin{tabular}{l c c c c c c c c}
\toprule
Model &  All & [0,50) & [50,100) & [100,200) & [200,300) & [300,400) & [400,500) & [500,$+\infty$) \\
\midrule
Base  & 84.326\% & \textbf{81.167}\% & 84.970\% & 83.023\% & 80.115\% & 79.378\% & 79.351\% & 74.829\% \\
MV-DNN  & 83.799\% & 81.094\% & 84.680\% & 82.641\% & 78.875\% & 79.470\% & 78.559\% & 73.939\% \\
MMDIN  & 84.810\% & 80.121\% & 84.322\% & 82.837\% & 80.371\% & 79.699\% & 80.394\% & 75.146\% \\
MIM & 84.512\% & 80.703\% & 84.817\% & 83.206\% & 80.814\% & 80.431\% & 80.973\% & 76.689\% \\
Savior & 84.926\% & 80.657\% & 84.949\% & 83.372\% & 81.057\% & 80.176\% & 81.179\% & 76.204\% \\
\textbf{PCR-CA} & \textbf{85.086}\% & 81.069\% & \textbf{85.050}\% & \textbf{83.499}\% & \textbf{81.266}\% & \textbf{80.691}\% & \textbf{81.505}\% & \textbf{76.977}\% \\
\midrule
\multicolumn{2}{l}{Improv} +0.760\% & -0.098\% & +0.080\% & +0.476\% & +1.151\% & +1.313\% & +2.154\% & +2.148\% \\
\bottomrule
\end{tabular}
\caption{Overall CTR prediction performance (AUC) across popularity buckets. ``All'' denotes the entire evaluation split (chronologically held-out test set). Buckets are defined by app popularity rank in store. Standard errors are below $0.01\%$ for all entries; all PCR-CA improvements over Base except [0,50) are statistically significant at $p<0.01$ (paired t-test).}
\label{tab:results}
\end{table*}

In this section, we compare PCR-CA with both models that leverage multimodal features and baseline models that do not. The overall performance is summarized in Table~\ref{tab:results}. Except for a slight decrease in the [0,50) bucket, PCR-CA consistently achieves substantial improvements in overall AUC and across all other buckets.

For head apps ([0,50) bucket), the Base model performs best. This is because top-ranked apps have abundant user interaction data, rendering multimodal features largely unnecessary; introducing them can slightly shift the model's attention and cause minor performance degradation. However, this decrease is negligible---PCR-CA drops by only 0.098\% compared to Base, which is insignificant relative to the overall gains.

\textbf{Head-app trade-off analysis.} The slight $-0.098\%$ AUC drop in the [0,50) bucket is an expected trade-off. Extremely popular head apps (e.g., \textit{Telegram}, \textit{Spotify}) already enjoy abundant interaction data, and users often visit the store with a clear intent to install them. In this regime, multimodal semantics provide little incremental signal, and shifting model capacity toward long-tail discovery marginally dilutes head-app precision. We view this as acceptable because our product goal is to help users discover previously unknown but relevant ``diamond'' apps; the overall business metric gains (Table~\ref{tab:abtest}) confirm that the long-tail improvement more than compensates for this negligible head-app degradation.

Notably, MV-DNN, which naively incorporates multimodal features, may result in performance degradation. This is due to the high dimensionality of multimodal features relative to ID features, which increases learning difficulty and introduces noise that outweighs the information gain. In contrast, MMDIN and MIM, which adopt more sophisticated multimodal structures, achieve further improvements. In our experiments, MMDIN slightly outperforms MIM, likely because the store item dataset is smaller than typical e-commerce datasets, and overly complex attention structures can impair performance---a trend also observed in our ablation study.

Savior, which employs a codebook to handle multimodal features, demonstrates clear performance gains, confirming the effectiveness of codebook-based representations. Nevertheless, due to the unique multi-category nature of apps, RQ-VAE is not optimal, whereas our parallel codebook design consistently yields superior results.

Furthermore, PCR-CA shows increasing effectiveness for long-tail apps, highlighting the critical role of multimodal features in recommending apps with sparse interaction data, where the model cannot learn effectively from user behavior alone.

\subsection{Ablation Studies}
We investigate the impact of various components in PCR-CA by removing or altering them.

\begin{table}[h]
\centering
\caption{Ablation: effect of contrastive alignment and codebook design, where ``w/o'' is short for ``without''. }
\label{tab:ablation-align-codebook}
\begin{tabular}{lcc}
\toprule
Model Variant & AUC (All) & vs PCR-CA \\
\midrule
w/o contrastive alignment & 84.936\% & -0.15\% \\
w/o codebook (raw content) & 84.337\% & -0.75\% \\
Single codebook (64-d VQ) & 84.600\% & -0.49\% \\
$K=1$ (one code per book) & 84.782\% & -0.30\% \\
Residual VQ (RQ-VAE) & 84.828\% & -0.26\% \\
\bottomrule
\end{tabular}
\end{table}

\begin{table}[h]
\centering
\caption{Ablation: effect of attention fusion strategies. }
\label{tab:ablation-attn}
\begin{tabular}{lcc}
\toprule
Fusion Variant & AUC (All) & vs PCR-CA \\
\midrule
ID-attention only (Base) & 85.042\% & -0.044\% \\
Cosine-similarity attention & 84.509\% & -0.577\% \\
MoE gating attention & 84.631\% & -0.455\% \\
All features in one attention & 85.054\% & -0.032\% \\
Multi-head attention & 85.119\% & +0.033\% \\
\bottomrule
\end{tabular}
\end{table}

Table~\ref{tab:ablation-align-codebook} shows the effect of removing contrastive alignment and modifying the codebook structure (evaluated by overall AUC). Table~\ref{tab:ablation-attn} compares different attention fusion strategies.

\textbf{Effect of Contrastive Alignment:} Removing $L_{\text{align}}$ (thus not explicitly aligning $\mathbf{s}$ and $\mathbf{v}$ spaces) drops overall AUC to 84.936\% (-0.15). This confirms that contrastive alignment helps the model better utilize the semantic embeddings. Without alignment, the attention tends to rely more on ID features, and the semantic part's contribution shrinks. With alignment, the content-based embedding of an item is pulled closer to users who like that item (via its ID embedding), making it more impactful.

\textbf{Effect of Codebook Quantization:} Using raw content (no quantization) yields 84.337\%, below PCR-CA (-0.75). So our quantization approach (with $N=4$ codebooks, etc.) consistently outperforms using a high-dimensional content vector directly, validating that the codebook is not just an efficiency hack but also improves the signal-to-noise ratio. A single codebook performs better than raw content (84.600\% vs 84.337\%) but worse than parallel (-0.49). It seems one codebook cannot capture all facets well; it might pick the dominant feature of an app but lose secondary features. Using a residual VQ (RQ-VAE) gave 84.828\%, close to SaviorRec's 84.926\%, since SaviorRec used a similar approach. This is still below our parallel codebooks. This supports our hypothesis: parallel coding is more effective for multi-aspect items than a residual coding. We also varied $K$. With $K=1$ (as in a standard VQ-VAE), AUC is 84.782\% (-0.30). So using a few codes per book ($K=2$) helps capture additional details, especially for items that combine multiple sub-aspects in one factor.

\begin{figure}[t]
\centering
\includegraphics[width=0.5\textwidth]{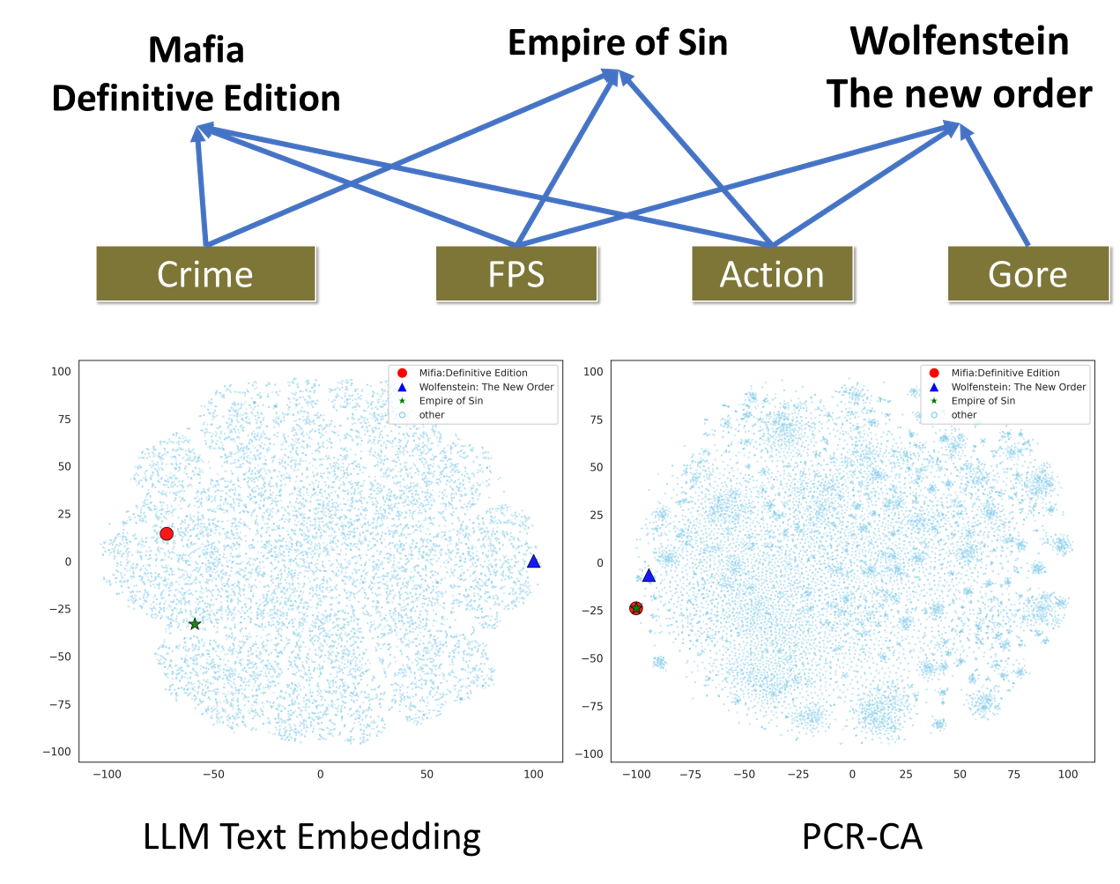}
\vspace{-2mm}
\caption{Visualization of multimodal embedding space. After applying PCR-CA, apps that share multiple categories exhibit closer semantic proximity, reflecting improved representation of multiple-category relationships.}
\label{fig:tsne}
\vspace{-4mm}
\end{figure}

Figure~\ref{fig:tsne} visualizes the semantic embedding space using t-SNE, with \textit{Mafia} as the anchor point. Ideally, \textit{Mafia} and \textit{Empire of Sin} should be placed close together, as both revolve around mafia narratives (crime) and share strong action/FPS elements. However, the raw LLM text model embeddings fail to sufficiently capture this similarity, leaving the two games apart. Meanwhile, \textit{Wolfenstein}, although not related to the mafia theme, also features intense action and FPS mechanics, yet its embedding is placed far from \textit{Mafia} as well. After applying our PCR-CA model, the embeddings shift considerably: \textit{Mafia} and \textit{Empire of Sin} collapse into near overlap, while \textit{Wolfenstein} also moves closer to this cluster. This demonstrates that our approach effectively integrates both thematic and gameplay elements into the learned representation, resulting in more accurate modeling of multiple-category relationships.

\textbf{Attention Fusion Strategies:} Table~\ref{tab:ablation-attn} compares ways of combining ID and semantic features in attention, on overall AUC. Using only ID-based attention gives 85.042\%. Replacing the learned attention MLP with a simple cosine similarity between $\mathbf{s}_c$ and each $\mathbf{s}_{i_j}$ (``cosine attention'') underperforms at 84.509\%. Using a learned MoE gate to weight the two attention modules (``MoE attention'') gives 84.631\%; it may overfit given the added complexity and small benefit. Concatenating ID and semantic features into a single attention network (``all features in one attention'') yields 85.054\%, just slightly below our dual attention 85.086\%. This suggests that a single attention can almost do the job if given both features, but dual attention provides a tiny edge. More importantly, we found dual attention made training more stable and it performed marginally better on tail subsets. Therefore, while the gain from dual attention over a simpler concatenation is small, we keep it as it consistently yielded the best results and ensures neither modality's signal is drowned out.

Although multi-head attention (MHA) improves AUC by 0.033\%, its parameter size ($\sim$10k vs. $\sim$1k for standard attention, with d=16) introduces substantial inference overhead on long sequences. Hence, we consider the marginal gain insufficient to justify the added complexity.
\subsection{Discussion on Diversity and Efficiency}
Beyond accuracy, we comment on two practical aspects:

\textbf{Diversity:} We measured the fraction of recommended apps (in top-10 lists for each user) that belong to the long tail (popularity rank >500). For the base model, only 12.4\% of recommendations were tail apps. For PCR-CA, this increased to 16.0\%. This indicates a notable shift towards more diverse recommendations. Of course, recommending more tail items is only good if those items are relevant; our AUC improvements show the model is indeed picking tail items that users engage with. By understanding item content, PCR-CA can match users to more unique or novel apps they might like, rather than always showing the same popular ones.

\textbf{Efficiency:} At serving time, PCR-CA requires storing an extra 64-d semantic embedding per item. But since we can store this as four small code indices, it's effectively storing 4 integers per item. The codebook vectors (4 * 32 * 16 floats) are negligible. So memory overhead is low. Inference involves an extra embedding lookup for the code embeddings and a slightly larger input to the MLP. In our simulation, the base model could score $\sim$150k candidate items per second by our server; PCR-CA scored $\sim$120k/sec on the same hardware---a moderate slowdown that we find acceptable. The bottleneck remained the number of items, not per-item complexity.

One added offline cost is computing text embeddings and code indices for new apps. But this is done once per app. We used a pre-trained LLM to embed descriptions. It cost 22 milliseconds per app. Thus, new apps can be onboarded in very short time, after which they're ready to recommend.

In summary, PCR-CA offers strong performance gains with minimal overhead, making it well-suited for production deployment---especially in enhancing long-tail content discovery without compromising recommendations for popular items.
\begin{table}[h]
\centering
\caption{Online A/B Test Results on Microsoft Store}
\label{tab:abtest}
\begin{tabular}{lcccc}
\hline
\textbf{Metric} & \textbf{Lift} & \textbf{95\% CI} & \textbf{p-value} & \textbf{Duration} \\
\hline
CTR & +10.52\% & [0.3413\%, 0.7838\%] & 0.03\% & 2 weeks \\
CVR & +16.30\% & [1.4261\%, 6.9407\%] & 0.29\% & 2 weeks \\
HER & +1.97\% & [0.15\%, 0.79\%] & 0.04\% & 2 weeks \\
\hline
\end{tabular}
\vspace{1mm}
\end{table}
\subsection*{Online A/B Test}
We deployed \textbf{PCR-CA} on the Microsoft Store homepage \textit{``Games/Apps we think you Love''} section and conducted a 2-week online A/B test covering \textit{50\% of total traffic} across four major markets, focusing on our core business objectives. Evaluation metrics included \textit{CTR}, \textit{CVR}, and \textit{HER (Highly Engaged Rate)}. The method achieved a \textbf{10.52\%} lift in CTR, \textbf{16.30\%} increase in CVR, and a \textbf{1.97\%} improvement in HER. Notably, the HER gain is specific to the homepage and limited by the collection's traffic share, rather than the experiment duration. All reported lifts are statistically significant (Table~\ref{tab:abtest}).

\begin{equation}
\text{HER} = \frac{\text{Engaged Users}}{\text{Total Users}}
\end{equation}

\vspace{1mm}
\noindent
\textit{Engaged Users} refers to users who performed active behaviors such as downloads or purchases. \textit{Total Users} refers to all users exposed to the recommendation module.

\section{Limitations and Future Work}
\label{sec:limitations}
Our work has several limitations. First, all experiments are conducted on proprietary Microsoft Store data, and due to company policy and privacy constraints, we cannot publicly release the dataset or code, which limits reproducibility. Second, we have not evaluated PCR-CA on other app stores (e.g., Google Play, iOS App Store) or public recommendation benchmarks; thus, generalization to other domains remains to be verified. Third, the algorithmic components---parallel VQ-AE, InfoNCE contrastive loss, and dual-stream attention---are individually standard; the primary contribution lies in their task-specific integration for multi-category app recommendation. Fourth, we deliberately excluded image features to avoid click-bait bias, which may sacrifice some visual signals that could benefit certain app categories. Finally, while our online A/B test demonstrates strong gains, the 2-week duration on 50\% traffic, though standard for industrial deployment, could be extended in future work to observe long-term user retention effects.

\section{Conclusion}
\label{sec:conclusion}
In this paper, we propose \textbf{PCR-CA}, a recommendation framework that integrates content and collaborative signals through parallel codebooks and contrastive alignment. PCR-CA learns multiple semantic codes per item to capture diverse attributes and aligns them with ID embeddings for behavior-aware representation. A dual-attention mechanism fuses ID-based and content-based signals efficiently, improving generalization without added complexity. Experiments on large-scale interaction logs from the Microsoft Store verify that PCR-CA outperforms strong baselines, demonstrating its effectiveness for long-tail and multi-category app recommendation in a real-world production environment.



\begin{thebibliography}{30}
\balance

\ifx \showCODEN    \undefined \def \showCODEN     #1{\unskip}     \fi
\ifx \showDOI      \undefined \def \showDOI       #1{#1}\fi
\ifx \showISSN     \undefined \def \showISSN      #1{\unskip}     \fi
\ifx \showLCCN     \undefined \def \showLCCN      #1{\unskip}     \fi
\ifx \shownote     \undefined \def \shownote      #1{#1}          \fi
\ifx \showarticletitle \undefined \def \showarticletitle #1{#1}   \fi
\ifx \showURL      \undefined \def \showURL       {\relax}        \fi
\providecommand\bibfield[2]{#2}
\providecommand\bibinfo[2]{#2}
\providecommand\natexlab[1]{#1}
\providecommand\showeprint[2][]{arXiv:#2}

\bibitem[Breiman(1996)]%
        {breiman1996bagging}
\bibfield{author}{\bibinfo{person}{Leo Breiman}.} \bibinfo{year}{1996}\natexlab{}.
\newblock \bibinfo{title}{\textit{Bagging Predictors}.}
\newblock
\newblock

\bibitem[Rendle(2010)]%
        {rendle2010fm}
\bibfield{author}{\bibinfo{person}{Steffen Rendle}.} \bibinfo{year}{2010}\natexlab{}.
\newblock \bibinfo{title}{\textit{Factorization Machines}.}
\newblock
\newblock

\bibitem[Oord et~al.(2017)]%
        {oord2017vqvae}
\bibfield{author}{\bibinfo{person}{Aaron van den Oord}, \bibinfo{person}{Oriol Vinyals}, {and} \bibinfo{person}{Koray Kavukcuoglu}.} \bibinfo{year}{2017}\natexlab{}.
\newblock \bibinfo{title}{\textit{Neural Discrete Representation Learning}.}
\newblock
\newblock

\bibitem[Oord et~al.(2018)]%
        {oord2018infonce}
\bibfield{author}{\bibinfo{person}{Aaron van den Oord}, \bibinfo{person}{Yazhe Li}, {and} \bibinfo{person}{Oriol Vinyals}.} \bibinfo{year}{2018}\natexlab{}.
\newblock \bibinfo{title}{\textit{Representation Learning with Contrastive Predictive Coding}.}
\newblock
\newblock

\bibitem[Elkahky et~al.(2015)]%
        {elkahky2015mv}
\bibfield{author}{\bibinfo{person}{Ali Elkahky}, \bibinfo{person}{Yang Song}, {and} \bibinfo{person}{Xiaodong He}.} \bibinfo{year}{2015}\natexlab{}.
\newblock \bibinfo{title}{\textit{A Multi-View Deep Learning Approach for Cross Domain User Modeling in Recommendation Systems}.}
\newblock
\newblock

\bibitem[Zheng et~al.(2025)]%
        {zheng2025semanticID}
\bibfield{author}{\bibinfo{person}{Wenwen Zheng}, \bibinfo{person}{Carolina Zheng}, {and} \bibinfo{person}{et al.}.} \bibinfo{year}{2025}\natexlab{}.
\newblock \bibinfo{title}{\textit{Enhancing Embedding Representation Stability in Recommendation Systems with Semantic ID}.}
\newblock
\newblock

\bibitem[Bahdanau et~al.(2015)]%
        {bahdanau2015neural}
\bibfield{author}{\bibinfo{person}{Dzmitry Bahdanau}, \bibinfo{person}{Kyunghyun Cho}, {and} \bibinfo{person}{Yoshua Bengio}.} \bibinfo{year}{2015}\natexlab{}.
\newblock \bibinfo{title}{\textit{Neural Machine Translation by Jointly Learning to Align and Translate}.}
\newblock
\newblock

\bibitem[Lin et~al.(2025)]%
        {lin2025unified}
\bibfield{author}{\bibinfo{person}{Guanyu Lin}, \bibinfo{person}{Julian McAuley}, {and} \bibinfo{person}{Stefano Ermon}.} \bibinfo{year}{2025}\natexlab{}.
\newblock \bibinfo{title}{\textit{Unified Semantic and ID Representation Learning for Deep Recommenders}.}
\newblock
\newblock

\bibitem[Zhou et~al.(2018)]%
        {zhou2018din}
\bibfield{author}{\bibinfo{person}{Guorui Zhou}, \bibinfo{person}{Xiaoqiang Zhu}, \bibinfo{person}{Chuanren Zhou}, {and} \bibinfo{person}{et al.}.} \bibinfo{year}{2018}\natexlab{}.
\newblock \bibinfo{title}{\textit{Deep Interest Network for Click-Through Rate Prediction}.}
\newblock
\newblock

\bibitem[Wang et~al.(2015)]%
        {wang2015cdl}
\bibfield{author}{\bibinfo{person}{Hao Wang}, \bibinfo{person}{Naiyan Wang}, {and} \bibinfo{person}{Dit-Yan Yeung}.} \bibinfo{year}{2015}\natexlab{}.
\newblock \bibinfo{title}{\textit{Collaborative Deep Learning for Recommender Systems}.}
\newblock
\newblock

\bibitem[He and McAuley(2016)]%
        {he2016vbpr}
\bibfield{author}{\bibinfo{person}{Ruining He} {and} \bibinfo{person}{Julian McAuley}.} \bibinfo{year}{2016}\natexlab{}.
\newblock \bibinfo{title}{\textit{VBPR: Visual Bayesian Personalized Ranking from Implicit Feedback}.}
\newblock
\newblock

\bibitem[Cheng et~al.(2016)]%
        {cheng2016wide}
\bibfield{author}{\bibinfo{person}{Heng-Tze Cheng}, \bibinfo{person}{Levent Koc}, \bibinfo{person}{Jeremiah Harmsen}, {and} \bibinfo{person}{et al.}.} \bibinfo{year}{2016}\natexlab{}.
\newblock \bibinfo{title}{\textit{Wide \& Deep Learning for Recommender Systems}.}
\newblock
\newblock

\bibitem[Guo et~al.(2017)]%
        {guo2017deepfm}
\bibfield{author}{\bibinfo{person}{Huifeng Guo}, \bibinfo{person}{Ruiming Tang}, \bibinfo{person}{Yunming Ye}, \bibinfo{person}{Zhenguo Li}, {and} \bibinfo{person}{Xiuqiang He}.} \bibinfo{year}{2017}\natexlab{}.
\newblock \bibinfo{title}{\textit{DeepFM: A Factorization-Machine based Neural Network for CTR Prediction}.}
\newblock
\newblock

\bibitem[Ma et~al.(2018)]%
        {ma2018mmoe}
\bibfield{author}{\bibinfo{person}{Jiaqi Ma}, \bibinfo{person}{Zhe Zhao}, \bibinfo{person}{Xinyang Yi}, \bibinfo{person}{Jilin Chen}, \bibinfo{person}{Lichan Hong}, {and} \bibinfo{person}{Ed H. Chi}.} \bibinfo{year}{2018}\natexlab{}.
\newblock \bibinfo{title}{\textit{Modeling Task Relationships in Multi-task Learning with Multi-gate Mixture-of-Experts}.}
\newblock
\newblock

\bibitem[Kang and McAuley(2018)]%
        {kang2018sasrec}
\bibfield{author}{\bibinfo{person}{Wang-Cheng Kang} {and} \bibinfo{person}{Julian McAuley}.} \bibinfo{year}{2018}\natexlab{}.
\newblock \bibinfo{title}{\textit{Self-Attentive Sequential Recommendation}.}
\newblock
\newblock

\bibitem[He et~al.(2021)]%
        {he2021hypergraph}
\bibfield{author}{\bibinfo{person}{Liang He}, \bibinfo{person}{Xiangnan He}, \bibinfo{person}{Fajie Yuan}, {and} \bibinfo{person}{et al.}.} \bibinfo{year}{2021}\natexlab{}.
\newblock \bibinfo{title}{\textit{Click-Through Rate Prediction with Multi-Modal Hypergraphs}.}
\newblock
\newblock

\bibitem[Yang et~al.(2021)]%
        {yang2021mmdin}
\bibfield{author}{\bibinfo{person}{Mingbao Yang}, \bibinfo{person}{Fajie Yuan}, \bibinfo{person}{Peng Wu}, {and} \bibinfo{person}{Xiao-Jun Wu}.} \bibinfo{year}{2021}\natexlab{}.
\newblock \bibinfo{title}{\textit{Multi-Head Multi-Modal Deep Interest Recommendation Network}.}
\newblock
\newblock

\bibitem[Zeghidour et~al.(2021)]%
        {zeghidour2021soundstream}
\bibfield{author}{\bibinfo{person}{Neil Zeghidour}, \bibinfo{person}{Alejandro Luebs}, \bibinfo{person}{Ahmed Omran}, \bibinfo{person}{Jan Skoglund}, {and} \bibinfo{person}{Marco Tagliasacchi}.} \bibinfo{year}{2021}\natexlab{}.
\newblock \bibinfo{title}{\textit{SoundStream: An End-to-End Neural Audio Codec}.}
\newblock
\newblock

\bibitem[Liu et~al.(2024)]%
        {liu2024tokenization}
\bibfield{author}{\bibinfo{person}{Qijiong Liu}, \bibinfo{person}{Zhijun Qin}, {and} \bibinfo{person}{Liang Zhang}.} \bibinfo{year}{2024}\natexlab{}.
\newblock \bibinfo{title}{\textit{Discrete Semantic Tokenization for Deep CTR Prediction}.}
\newblock
\newblock

\bibitem[Wang et~al.(2017)]%
        {wang2017deep}
\bibfield{author}{\bibinfo{person}{Ruoxi Wang}, \bibinfo{person}{Bin Fu}, \bibinfo{person}{Gang Fu}, {and} \bibinfo{person}{Mingliang Wang}.} \bibinfo{year}{2017}\natexlab{}.
\newblock \bibinfo{title}{\textit{Deep \& Cross Network for Ad Click Predictions}.}
\newblock
\newblock

\bibitem[Salakhutdinov and Hinton(2009)]%
        {salakhutdinov2009semantic}
\bibfield{author}{\bibinfo{person}{Ruslan Salakhutdinov} {and} \bibinfo{person}{Geoffrey Hinton}.} \bibinfo{year}{2009}\natexlab{}.
\newblock \bibinfo{title}{\textit{Semantic Hashing}.}
\newblock
\newblock

\bibitem[Zhang et~al.(2019)]%
        {zhang2019survey}
\bibfield{author}{\bibinfo{person}{Shuai Zhang}, \bibinfo{person}{Lina Yao}, \bibinfo{person}{Aixin Sun}, {and} \bibinfo{person}{Yi Tay}.} \bibinfo{year}{2019}\natexlab{}.
\newblock \bibinfo{title}{\textit{Deep Learning Based Recommender System: A Survey and New Perspectives}.}
\newblock
\newblock

\bibitem[Song et~al.(2019)]%
        {song2019autoint}
\bibfield{author}{\bibinfo{person}{Weiping Song}, \bibinfo{person}{Chence Shi}, \bibinfo{person}{Zhiping Xiao}, \bibinfo{person}{Zhijian Duan}, \bibinfo{person}{Yewen Xu}, \bibinfo{person}{Ming Zhang}, {and} \bibinfo{person}{Jian Tang}.} \bibinfo{year}{2019}\natexlab{}.
\newblock \bibinfo{title}{\textit{AutoInt: Automatic Feature Interaction Learning via Self-Attentive Neural Networks}.}
\newblock
\newblock

\bibitem[Zhao et~al.(2023)]%
        {zhao2023bootstrapping}
\bibfield{author}{\bibinfo{person}{Xinping Zhao}, \bibinfo{person}{Ying Zhang}, \bibinfo{person}{Qiang Xiao}, {and} \bibinfo{person}{Yingchun Yang}.} \bibinfo{year}{2023}\natexlab{}.
\newblock \bibinfo{title}{\textit{Bootstrapping Contrastive Learning Enhanced Music Cold-Start Matching}.}
\newblock
\newblock

\bibitem[Yan et~al.(2025)]%
        {yan2025mim}
\bibfield{author}{\bibinfo{person}{Bencheng Yan}, \bibinfo{person}{Si Chen}, \bibinfo{person}{Shichang Jia}, \bibinfo{person}{Jianyu Liu}, \bibinfo{person}{Yueran Liu}, \bibinfo{person}{Chenghan Fu}, \bibinfo{person}{Wanxian Guan}, \bibinfo{person}{Hui Zhao}, \bibinfo{person}{Xiang Zhang}, \bibinfo{person}{Kai Zhang}, \bibinfo{person}{Wenbo Su}, \bibinfo{person}{Pengjie Wang}, \bibinfo{person}{Jian Xu}, \bibinfo{person}{Bo Zheng}, {and} \bibinfo{person}{Baolin Liu}.} \bibinfo{year}{2025}\natexlab{}.
\newblock \bibinfo{title}{\textit{MIM: Multi-modal Content Interest Modeling Paradigm for User Behavior Modeling}}.
\newblock
\newblock
\newblock
\shownote{arXiv preprint arXiv:2502.00321}.

\bibitem[Deldjoo et~al.(2020)]%
        {deldjoo2020multimedia}
\bibfield{author}{\bibinfo{person}{Yashar Deldjoo}, \bibinfo{person}{Mohamed Trabelsi}, {and} \bibinfo{person}{Hamed Zamani}.} \bibinfo{year}{2020}\natexlab{}.
\newblock \bibinfo{title}{\textit{Recommender Systems Leveraging Multimedia Content}.}
\newblock
\newblock

\bibitem[Yao et~al.(2025)]%
        {yao2025saviorrec}
\bibfield{author}{\bibinfo{person}{Yining Yao}, \bibinfo{person}{Yang Zhao}, \bibinfo{person}{Zhuoye Ding}, {and} \bibinfo{person}{Jian Xu}.} \bibinfo{year}{2025}\natexlab{}.
\newblock \bibinfo{title}{\textit{SaviorRec: Semantic-Behavior Alignment for Cold-Start Recommendation}.}
\newblock
\newblock

\bibitem[Wei et~al.(2021)]%
        {wei2021contrastive}
\bibfield{author}{\bibinfo{person}{Yinwei Wei}, \bibinfo{person}{Xiang Wang}, \bibinfo{person}{Qi Li}, \bibinfo{person}{Liqiang Nie}, \bibinfo{person}{Yan Li}, \bibinfo{person}{Xuanping Li}, {and} \bibinfo{person}{Tat-Seng Chua}.} \bibinfo{year}{2021}\natexlab{}.
\newblock \bibinfo{title}{\textit{Contrastive Learning for Cold-Start Recommendation}.}
\newblock
\newblock

\bibitem[Zhu et~al.(2024)]%
        {zhu2024cost}
\bibfield{author}{\bibinfo{person}{Yongchao Zhu}, \bibinfo{person}{Xiangyu Zhao}, {and} \bibinfo{person}{Fajie Yuan}.} \bibinfo{year}{2024}\natexlab{}.
\newblock \bibinfo{title}{\textit{CoST: Content Semantic Tokenization for Generative Recommendation}.}
\newblock
\newblock

\bibitem[Pan et~al.(2021)]%
        {pan2021click}
\bibfield{author}{\bibinfo{person}{Yujie Pan}, \bibinfo{person}{Jiangchao Yao}, \bibinfo{person}{Bo Han}, \bibinfo{person}{Kunyang Jia}, \bibinfo{person}{Ya Zhang}, {and} \bibinfo{person}{Hongxia Yang}.} \bibinfo{year}{2021}\natexlab{}.
\newblock \bibinfo{title}{\textit{Click-through Rate Prediction with Auto-Quantized Contrastive Learning}.}
\newblock
\newblock

\bibitem[Jegou et~al.(2011)]%
        {jegou2011product}
\bibfield{author}{\bibinfo{person}{Herve Jegou}, \bibinfo{person}{Matthijs Douze}, {and} \bibinfo{person}{Cordelia Schmid}.} \bibinfo{year}{2011}\natexlab{}.
\newblock \bibinfo{title}{\textit{Product Quantization for Nearest Neighbor Search}.}
\newblock
\newblock

\end{thebibliography}
\end{document}